\documentclass[preprint,pre,showpacs,amsmath]{revtex4}
\usepackage{graphicx}
\begin{document}

\title{A Thermodynamic Model of Electric-Field-Induced Pattern Formation in
Binary Dielectric Fluids}

\author{M.D. Johnson}
\email{mjohnson@ucf.edu}

\author{X. Duan}
\author{Brett Riley}

\author{Aniket Bhattacharya}
\email{aniket@physics.ucf.edu}

\author{Weili Luo}
\email{luo@ucf.edu}

\affiliation{Department of Physics, University of Central Florida, Orlando, FL
32816-2385}

\date{November 20, 2002}

\begin{abstract}
An electric-field-induced phase transition and pattern formation
in a binary dielectric fluid layer are studied using a coarse-grained
free energy functional. The electrostatic part of the free energy is a
nonlinear functional of the dielectric function, which depends in turn
on the local colloidal concentration.  We determine the phase
co-existence curve and find that beyond a critical electric field the
system phase separates.  Accompanying the phase separation are patterns
similar to those observed in a spinodal decomposition of an ordinary
binary fluid. The temporal evolution of the phase separating patterns
are discussed both analytically and numerically by integrating a
Cahn-Hilliard type of equation.
\end{abstract}


\maketitle

\section{Introduction}

Systems undergoing a phase separation from a disordered phase to a more
ordered state often form interesting patterns as they select and
approach the final state\cite{Cross}. The process of phase ordering, 
or coarsening dynamics, can be classified into different universality classes 
depending on the physical dimension, the dimension of the order parameter,
and conservation laws\cite{Gunton,Bray}. Typically the associated
scaling properties and dynamical universalities
can be understood using a coarse-grained free energy functional. A 
time-dependent Ginzburg-Landau approach then allows study of the
evolving patterns.

Ferro-fluids\cite{Ferro}, which are colloidal suspensions of
magnetic particles in solution, exhibit complex labyrinthine 
pattern formation when trapped between closely spaced glass plates
and subjected to a transverse magnetic field\cite{Ferro,Langer,wang}. 
The complex
patterns result from competition between the surface tension and  
long-range forces, and have much in common with patterns observed in
amphiphilic systems and Langmuir-Blodget films. In this paper we report
on a similar effect in colloidal suspensions in which phase separation
and a concomitant pattern formation are driven by an external electric
field.  Our results are based upon a macroscopic thermodynamic
model of a dielectric binary liquid.  This work was motivated by recent
experiments demonstrating such behavior\cite{experiment1, experiment2},
and shares some features in common with the
experiments. We reported some preliminary results in
Ref.~\onlinecite{conference}.

We consider colloids consisting of nanometer scale (typical diameter
$\sim 10\,$nm)
dielectric particles suspended in a dielectric solution. In the present
work we are interested in patterns on length scales much larger than the
particle size and the typical inter-particle spacing. Consequently we
use a coarse-grained description similar to the model B  used to study
phase separation in binary liquids\cite{Gunton,Hohenberg}. For an order parameter
we use the local particle concentration or volume fraction 
$c(\mathbf{r})$ ($0\le c \le 1$).  We model the particle
interaction using a local dielectric constant
$\varepsilon(c(\mathbf{r}))$ which depends nonlinearly on the local
concentration.  The use of a phenomenological dielectric constant
with a positive curvature is an
important ingredient of the model. It turns out that beyond a critical
value of the electric field the electrostatic energy dominates and
drives phase separation and pattern formation.  

The approach developed in this paper is thermodynamic, based on
a macroscopic free energy. Analyzing equilibrium properties
yields a phase diagram which exhibits instabilities that can lead
to pattern formation. Some understanding of the dynamics of
pattern formation then comes from investigating the corresponding
time dependent Ginzburg-Landau equation. We begin by describing the model free energy
and its equilibrium properties, and then turn to dynamics.

\section{Model Free Energy}

Consider a binary dielectric liquid as described above placed between parallel
capacitor plates separated by $d$. The liquid has capacitance
\begin{equation}
C = \frac{1}{4\pi d^2} \int d^3 r \, \varepsilon(c(\mathbf{r})) .
\label{C}
\end{equation}
When the internal energy is dominated by the electrostatic contribution,
the Helmholtz free energy is
\begin{equation}
F = \frac{1}{2}QV - TS,
\label{F}
\end{equation}
where $V$ is the voltage between the
plates, $T$ the temperature, and $S$ the entropy. The equilibrium state
is obtained by minimizing $F$ at constant $Q,T$. For our purposes it is
more convenient to make a Legendre transformation to a different free energy:
\begin{equation}
\Omega = F - QV = -\frac{1}{2}QV - TS .
\label{Omega1}
\end{equation}
Equilibrium is found by minimizing $\Omega$ at constant $V,T$ (note that
$V$ is voltage and not volume). We will refer to $\Omega$ as ``the''
free energy.

For the dielectric constant $\varepsilon$ we use a phenomenological model
due to Lichteneker\cite{dielectric, experiment2}:
\begin{equation}
\varepsilon(c) = \varepsilon_f e^{\gamma c}, \mbox{\ where\ }
\gamma = \ln \varepsilon_p / \varepsilon_f .
\label{epsilon}
\end{equation}
This interpolates between the dielectric constant of
pure solvent ($\varepsilon_f$) and that of pure particle ($\varepsilon_p$)
as the concentration varies from $c=0$ to $c=1$. In this paper we
use a physically reasonable value $\gamma=3$\cite{experiment1, experiment2}.
The specific choice of $\varepsilon(c)$ in Eq.~(\ref{epsilon}) is not
essential. The results
we report turn out to be largely independent of the form of
$\varepsilon(c)$, as long as it has positive curvature (a point
we discuss later).

We use an approximate entropy functional appropriate for interacting
particles at low concentration \cite{experiment1,experiment2,sear,bibette}:
\begin{equation}
S = - \frac{k_B}{v} \int d^3 r \big[ c(\mathbf{r}) \ln c(\mathbf{r})
+ (1-c(\mathbf{r})) \ln (1 - c(\mathbf{r})) \big] .
\label{S}
\end{equation}
This expression can be obtained by a simple counting argument.
In the absence of interactions $v$ is the particle volume and
Eq.~(\ref{S}) is the entropy of free particles \cite{Kittel}.
Here however $v$ is instead the so-called `correlation volume,' which
accounts phenomenologically for the particle interactions. 

In a typical experiment the
plate separation $d$ is so small that the concentration $c(\mathbf{r})$
can be treated as effectively two-dimensional. Then we can let the coordinate
$\mathbf{r}=(x,y)$ denote lateral position within the plane. In this limit
the electric field $E$ becomes uniform ($E=V/d$),
and the free energy becomes
\begin{subequations}
\label{Omega}
\begin{eqnarray}
\Omega &=& d \int d^2 r\, f(c(\mathbf{r})), \quad \mbox{where}\\
\label{fc}
f(c) &=& - \frac{E^2}{8\pi}\varepsilon(c) + \frac{k_BT}{v}
[ c \ln c + (1-c) \ln(1-c) ].
\end{eqnarray}
\end{subequations}
For $\varepsilon(c)$ given in Eq.~(\ref{epsilon}),
the temperature and electric field naturally scale to the
dimensionless form
\begin{equation}
\Tilde{T} = \frac{k_B T/v}{\varepsilon_f E^2/ 8\pi}.
\label{t}
\end{equation}

\section{Equilibrium Phase Separation}

\begin{figure}
\begin{center}
\includegraphics*[scale=0.35]{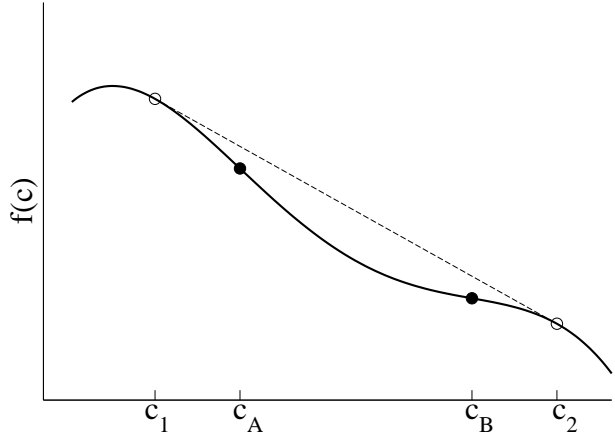}
\end{center}
\caption{Two points on a curve can have a common tangent
only if the curve has at least two inflection points. The inflection points
at $c_A,c_B$ are indicated by filled circles and the phase separation
points at $c_1,c_2$ by open circles.}
\label{inflection}
\end{figure}

We are interested in potential phase separations in the system. 
These can occur if $f(c)$ possesses inflection points\cite{Plischke}.
Suppose that in equilibrium the system phase separates into
volume fractions $x_{1,2}$ with concentrations $c_{1,2}$.
If the average concentration is $c$, then these must satisfy
\begin{equation}
\sum_{i=1}^2 x_i=1, \quad \sum_{i=1}^2 x_i c_i = c,
\label{constraint}
\end{equation}
and so
\begin{equation}
x_1 = \frac{c_2-c}{c_2-c_1}, \quad x_2 = \frac{c-c_1}{c_2-c_1}.
\label{x}
\end{equation}
To find the conditions under which the equilibrium state is
phase-separated in this manner, we minimize the free energy
$\sum_i x_i f(c_i)$ with respect to $c_1,c_2,x_1,x_2$, subject
to the constraints Eq.~(\ref{constraint}).
Using Lagrange multipliers, this means minimizing
\begin{equation}
\tilde{\Omega} = \sum_{i=1}^2 x_i [ f(c_i) - \mu c_i - \lambda] .
\label{tomega}
\end{equation}
Except at the endpoints ($x_i$ or $c_i$ equal to 0 or 1) this minimization
yields
\begin{subequations}
\label{f}
\begin{eqnarray}
\label{fa}
f(c_i) &=& \mu c_i + \lambda  \\
\label{fp}
f'(c_i) &=& \mu .
\end{eqnarray}
\end{subequations}
If $c_1$ and $c_2$ both satisfy Eq.~(\ref{fa}), then
\begin{equation}
f(c_2)-f(c_1) = \mu (c_2 - c_1) .
\label{line}
\end{equation}
This result is summarized by a simple geometrical construction illustrated
in Fig.~\ref{inflection}.  If the straight line joining the points
$(c_1,f(c_1))$ and $(c_2,f(c_2))$ is tangent to $f(c)$ at both points,
then phase separation will occur for $c_1\le c \le c_2$. This can occur
only if $f(c)$ has two or more inflection points (see Fig.~\ref{inflection}).
For our model, the endpoints $c=0,1$ never correspond to a minimum
free-energy state.

\begin{figure}
\begin{center}
\includegraphics*[scale=0.35]{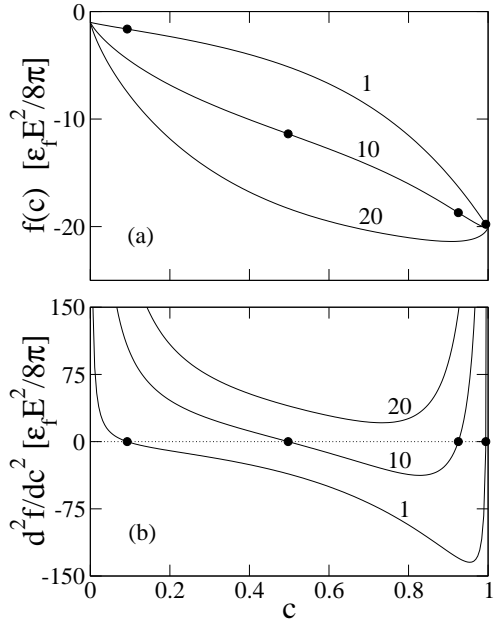}
\end{center}
\caption{(a) $f(c)$ \textit{vs}. $c$ for the three scaled temperatures 
$\Tilde{T}=1, 10, 20$, using Eqs.~(\ref{fc},\ref{epsilon}) with $\gamma=3$.  
At the two lower temperatures two inflection points are present.  
These are more easily seen in (b) $f''(c)$ \text{vs}. $c$. The
inflection points are indicated by filled circles.}
\label{fplot}
\end{figure}

Consider now the free energy density $f(c)$ for our dielectric fluid,
Eq.~(\ref{fc}). The entropic term $-TS$ has positive curvature. At
high temperatures this term dominates, $f(c)$ has positive curvature
everywhere,  and the equilibrium state is homogeneous. This is illustrated
by the highest-temperature trace ($\tilde{T}=20$) in Fig.~\ref{fplot},
which uses the model dielectric function Eq.~(\ref{epsilon}).
As the temperature is lowered the electrostatic energy term
$-QV/2=-E^2\varepsilon(c)/8\pi$ plays an increasingly important role. 
When $\varepsilon(c)$ has positive curvature, the electrostatic
term has negative curvature,
and at low temperatures this term leads to inflection points and
phase separation. This is illustrated by the two lower-temperature traces
in Fig.~\ref{fplot}.

\begin{figure}
\begin{center}
\includegraphics*[scale=0.4]{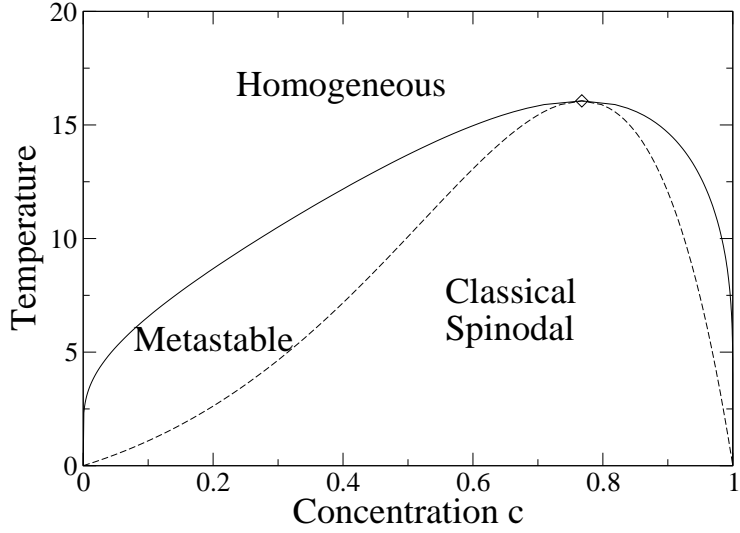}
\end{center}
\caption{Phase diagram for the free energy given in 
Eqs.~(\ref{fc},\ref{epsilon}) for $\gamma=3$. Temperature is $\tilde{T}$.
Shown are the coexistence curve
$\tilde{T}_{\mathrm{coex}}$ (solid) and the spinodal $\tilde{T}_s$ (dashed).
The critical point Eq.~(\ref{crit}) is marked by a diamond.}
\label{phase}
\end{figure}

The phase diagram for the free energy Eq.~(\ref{fc}), developed
using the above procedure, is shown in
Fig.~\ref{phase} for $\gamma=3$.  Below the coexistence curve
(the solid line) the homogeneous phase becomes metastable and in
equilibrium the system phase-separates. This curve is the locus of
points $c_{1,2}$ obtained from solving Eq.~(\ref{f}) at all 
temperatures. Although there is no exact expression for the
coexistence temperature, for the physically important case of low
concentrations we find approximately
\begin{equation}
\tilde{T}_{\mathrm{coex}} \approx \frac{(1+\gamma - \gamma c)e^{\gamma c}-
e^{\gamma}}{\ln c}.
\label{Tcoex}
\end{equation}
For $\gamma=3$ this is accurate for concentrations $c\lesssim 0.3$.

Lying below the dashed phase boundary in Fig.~\ref{phase} is the
classical spinodal region, where the homogeneous state becomes
unstable [$f''(c)<0$]. This phase boundary is given by the condition $f''(c)=0$,
which yields
\begin{equation}
\tilde{T}_s = \gamma^2 e^{\gamma c} c(1-c) .
\label{Ts}
\end{equation}

The coexistence and spinodal curves touch at a critical point
(marked by a diamond in Fig.~\ref{phase}). This occurs at the
critical concentration $c_c$ and critical temperature $T_c$ where
the two inflection points present at lower temperatures coincide
[and so $f''(c_c)=0=f'''(c_c)$].  For the model
dielectric function Eq.~(\ref{epsilon}) the critical values are:
\begin{subequations}
\label{crit}
\begin{eqnarray}
c_c &=& \frac{1}{2\gamma}\left(\gamma-2+\sqrt{\gamma^2+4}\right)\\
\Tilde{T}_c &=& \gamma^2 e^{\gamma c_c} c_c(1-c_c) .
\end{eqnarray}
\end{subequations}
The instabilities give the possibility of pattern formation,
which we now discuss.

\section{Pattern Formation: Dynamics}

Pattern formation after quenches into the metastable or spinodal regions
provides important insight into the underlying mechanisms for phase
transformation in the system. For a system like the one modeled here
it is experimentally easier to increase the voltage suddenly while
holding the temperature fixed; by Eq.~(\ref{t}) this is equivalent
to a quench.  Here we investigate pattern formation
under quenches using a phenomenological diffusion equation developed
from the free energy.

Notice first that when inhomogeneities arise the free energy needs
an additional $|\nabla c|^2$ term. This can be viewed as the surface
energy between domains of different concentration; such
terms arise generally in any interacting system. Consequently
in this section we write the free energy
\begin{equation}
\Omega = \int d^3r\, \left[ f(c(\mathbf{r})) + \frac{1}{2} \kappa 
\left| \nabla c  \right|^2 \right],
\label{OmegaK}
\end{equation}
where $\kappa$ is a phenomenological parameter.

Fick's law relates the concentration current density to the local chemical
potential $\mu$:
\begin{equation}
\mathbf{j}(\mathbf{r}) = -Mv_0 \nabla \mu(\mathbf{r}),
\label{j}
\end{equation}
where $M$ is the mobility (taken to be constant) and $v_0$ is the
volume of the colloidal particle.
Combining this with the continuity equation
$\dot c + \nabla\cdot\mathbf{j}=0$ yields a
diffusion equation
\begin{equation}
\frac{\partial c}{\partial t} = Mv_0 \nabla^2 \mu .
\label{de}
\end{equation}
The local chemical potential is obtained from the free energy:
\begin{equation}
\mu(\mathbf{r}) = v_0 
\left.\frac{\delta \Omega}{\delta c(\mathbf{r})}\right|_{TE}
= v_0 \left[ f'(c(\mathbf{r})) - \kappa \nabla^2 c(\mathbf{r}) \right].
\label{mu}
\end{equation}
Inserting this into Eq.~(\ref{de}) gives 
\begin{equation}
\frac{\partial c}{\partial t} =  Mv_0^2 \left[
\nabla^2 f'(c(\mathbf{r})) - \kappa \nabla^4 c(\mathbf{r}) \right].
\label{tdgl}
\end{equation}
The free energy density $f(c)$ is in Eqs.~(\ref{fc},\ref{epsilon}).
A convenient dimensionless form comes from writing time in units of
$\tau$ and length in units of $\lambda$, where
\begin{subequations}
\label{units}
\begin{eqnarray}
\tau &=& \frac{\kappa}{Mv_0^2}\left(\frac{8\pi}{\varepsilon_f E^2}\right)^2\\
\lambda &=& \sqrt{\frac{8\pi\kappa}{\varepsilon_f E^2}}.
\end{eqnarray}
\end{subequations}
\begin{subequations}
\label{diffusion}
Then the diffusion equation becomes
\begin{equation}
\frac{\partial c}{\partial t} =  \nabla^2 \left[ \tilde{f}'(c) 
    - \nabla^2 c \right],
\label{gleqn}
\end{equation}
where
\begin{equation}
\tilde{f}(c) = -e^{\gamma c} +
    \tilde{T} \left[ c\ln{c} + (1-c)\ln(1-c)\right].
\end{equation}
\end{subequations}
Eq.~(\ref{gleqn}) is nothing but the time-dependent Ginzburg-Landau
equation for a scalar conserved order parameter; however, the free
energy in this particular case is constructed with the parameters
directly relevant to experiments on dielectric colloidal 
fluids\cite{experiment1,experiment2}.
We study pattern formation after quenches by
solving this nonlinear diffusion equation, starting
from a uniform distribution (plus a small random component).

One can get some intuition about dynamics by linearizing the
diffusion equation\cite{Gunton}. The result is accurate for early times, before
the nonlinear terms become significant.
Write $c(\mathbf{r}) = c + \delta c(\mathbf{r})$,
where $c$ is the average concentration. For small $\delta c$ expand
\begin{equation}
\tilde{f}(c+\delta c) = \tilde{f}(c) + \tilde{\mu}_0 \delta c
- \frac{1}{2}b\left(\delta c\right)^2 + \dots,
\label{fseries}
\end{equation}
where
\begin{subequations}
\begin{eqnarray}
\tilde{\mu}_0 &=& \tilde{f}'(c) = -\gamma e^{\gamma c} + \tilde{T}
[\ln{c}-\ln(1-c)]\\
\label{b}
b &=& -\tilde{f}''(c) = \frac{\tilde{T}_s-\tilde{T}}{c(1-c)} .
\end{eqnarray}
\end{subequations}
$\tilde{T}_s$ is given in Eq.~(\ref{Ts}).  The linear and constant
terms in Eq.~(\ref{fseries}) play no role in the dynamics. Keeping only
the second order term from Eq.~(\ref{fseries}),
the free energy Eq.~(\ref{OmegaK}) becomes
\begin{equation}
\Omega \approx \frac{1}{2} \int d^3r\, \left[ \kappa |\nabla \delta c|^2 
 - b (\delta c)^2 \right].
\label{OmegaApprox}
\end{equation}

This expansion connects our model to the coarsening dynamics of a binary
alloy described by a conserved scalar order parameter $\phi(\mathbf{r})$.
The free energy for such a system is generally assumed to have the
Ginzburg-Landau form
\begin{equation}
F = \frac{1}{2}{\int}d^3r\,[{\kappa}|{\nabla}\phi|^2
 - b\phi^2 + \frac{u}{2}\phi^4] 
\label{bf}
\end{equation}
where $\kappa$, and $u$ are assumed to be positive. Below the critical
temperature $b$ becomes positive, yielding a broken symmetry.
Higher-order terms beyond $\phi^4$ are irrelevant for the dynamic
universality class. In fact, the early-time dynamics are dominated by
$\phi^2$ terms.  From the correspondence between
Eqs.~(\ref{OmegaApprox}), (\ref{bf}) we expect our model to have
early-time dynamics very similar to those of model B.

Using the expansion Eq.~(\ref{OmegaApprox})
linearizes the diffusion equation Eq.~(\ref{diffusion}):
\begin{equation}
\frac{\partial (\delta c)}{\partial t} = -\nabla^2 \left[ b\, \delta c + \nabla^2
\delta c \right].
\end{equation}
This is solved by expanding the fluctuation $\delta c$ in Fourier components:
\begin{equation}
\delta c(\mathbf{r},t) = \int \frac{d^2k}{(2\pi)^2}
e^{i\mathbf{k}\cdot\mathbf{r}} c(\mathbf{k},t) .
\end{equation}
We find
\begin{equation}
c(\mathbf{k},t) = c(\mathbf{k},0) e^{\alpha_k t},\mbox{\ where\ }
\alpha_k = k^2 (b-k^2).
\end{equation}
Outside of the spinodal region $\tilde{T}>\tilde{T}_s$ and so by
Eq.~(\ref{b}) $b<0$. Consequently all components $c(\mathbf{k},t)$ decay.
This describes the homogeneous equilibrium state: fluctuations away from
a uniform concentration decay.

However, within the classical spinodal region $b>0$. Thus
long wavelength modes, those with wave vector $k<\sqrt{b}$, grow
exponentially.
The most rapid growth occurs for $k_m=\sqrt{b/2}$, where $\alpha_k$ reaches
its maximum value $b^2/4$.
Thus in the classical spinodal regime, where 
the homogeneous state is unstable, early-time exponential growth leads to
structures a typical (dimensionless) length scale
$k_m^{-1}\sim (\tilde{T}_s-T)^{-1/2}$. 
These structures grow exponentially in time with a (dimensionless)
time scale $1/\alpha_{k_m}=4/b^2\sim(\tilde{T}_s-\tilde{T})^{-2}$. 

For example, consider the structure factor
\begin{equation}
S(\mathbf{k},t) = \langle c(\mathbf{k},t) c(-\mathbf{k},t) \rangle .
\end{equation}
Here angle brackets represent a statistical average.
Suppose that the initial fluctuations $c(\mathbf{k},0)$ are
small and uncorrelated: 
\begin{equation}
\langle c(\mathbf{k},0) c(\mathbf{k}',0) \rangle = A^2 \delta^2(\mathbf{q}
-\mathbf{q}') .
\end{equation}
Then for early times
\begin{equation}
S(\mathbf{k},t) = A^2 e^{2 \alpha_k t} .
\end{equation}
This grows ever more peaked at $k_m$ as time increases. Fourier
transforming gives the real-space correlation function
\begin{equation}
\langle \delta c(\mathbf{r},t) \delta c(\mathbf{r}',t) \rangle
\sim \frac{A^2}{t^{1/2}} e^{b^2 t/4} J_0(k_m|\mathbf{r}-\mathbf{r}'|) 
\end{equation}
which exhibits the length and time scales described above.

Including dimensions, the structures developing in early times
have a length scale
\begin{equation}
L \sim \frac{\lambda}{k_m} = \sqrt{\frac{8\pi\kappa}{\varepsilon_f E^2}}
\sqrt{\frac{2c(1-c)}{\tilde{T}_s-\tilde{T}}}= 
\sqrt{\frac{2c(1-c)k_B/v}{T_s-T}}
\label{L}
\end{equation}
and grow with a time scale
\begin{equation}
t_0 \sim \frac{4\tau}{b^2} = 
\frac{\kappa}{Mv_0^2}\left(\frac{8\pi}{\varepsilon_f E^2}\right)^2 
\left[ \frac{2c(1-c)}{\tilde{T}-\tilde{T}_s} \right]^2 =
\frac{\kappa}{Mv_0^2} \left[ \frac{2c(1-c)k_B/v}{T-T_s} \right]^2 .
\end{equation}

Deep quenches into the classical spinodal region should lead 
to dynamically-developing structures on a length scale $L$ given by 
Eq.~(\ref{L}). These structures result from a long-wavelength instability.
They represent not the equilibrium
state, but the early-time evolution from the highly non-equilibrium 
homogeneous state toward the (eventual) phase-separated equilibrium.
The linearized description is accurate for very small initial times.
Soon the fluctuations grow large enough for the nonlinear terms to
play a role, and we turn to numerical means for the solution of 
Eq.~(\ref{diffusion}).

\section{Numerical Integration of the Diffusion Equation}

We numerically integrate the lattice discretized version of 
Eqs.~(\ref{diffusion}) using a first-order Euler scheme, choosing
step sizes to avoid unphysical instabilities\cite{Gunton2}.
This algorithm is adequate for times during which the
system begins to organize into patterns. To carry the simulation
all the way to phase separation requires a more sophisticated
approach\cite{Gunton3}. However the model we use also omits
hydrodynamic effects which may be important for the late time 
coarsening process. Consequently here we present results
during the era of pattern formation only.

The initial configuration consists of a uniform concentration
$c$ plus a small random component. Starting from this initial
state, Eq.~(\ref{diffusion}) is integrated for various values 
of $\gamma$ and $\tilde{T}$.
In agreement with the results of the previous 
sections we find that phase separation occurs below the
coexistence temperature $\tilde{T}_{\mathrm{coex}}$
(or, equivalently, above a threshold electric field).
In the phase-separation region
the concentration develops labyrinthine patterns consisting of
stripes of higher and lower concentration.

\begin{figure}
\begin{center}
\includegraphics*[scale=0.4,angle=270]{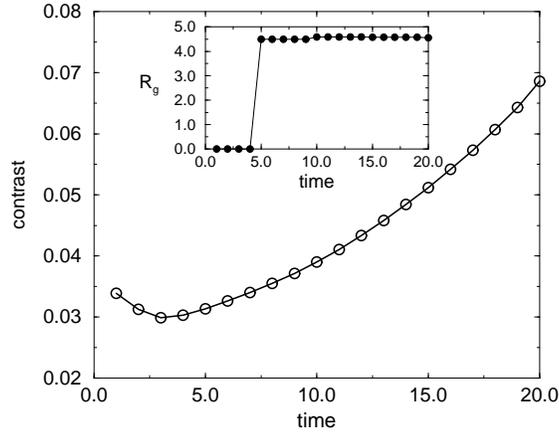}
\end{center}
\caption{Early stage increase in contrast
$(c_{\mathrm{max}}-c_{\mathrm{min}})/2c$ and
domain size (inset) as a function of time. Here time and space are in
the units of Eq.~(\ref{units}).
The simulation was done on a 128$\times$128 lattice with an initial random 
configuration $\pm 0.01$ around $c = 0.1$ for $\gamma = 1.0$ and 
$\Tilde{T} = 0.045$.}
\label{contrast}
\end{figure}

\begin{figure}
\begin{center}
\includegraphics*[scale=0.40,angle=270]{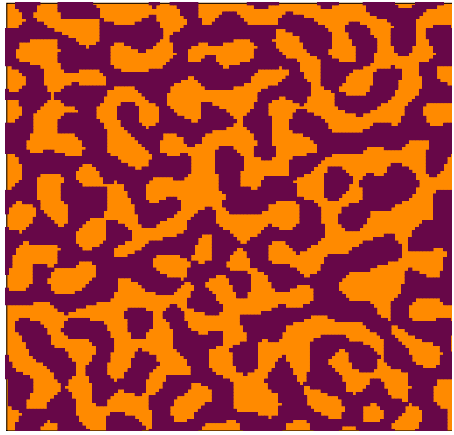}
\end{center}
\caption{A typical snapshot at time 20. Concentrations greater (less) than
the average $c$ are dark (light).}
\label{snapshot}
\end{figure}

Typical results for a quench deep into the phase separation region
are shown in Figs.~\ref{contrast} and \ref{snapshot}.
The latter clearly shows
that a labyrinthine pattern has developed.  

We monitor two quantities during the simulation. At each time
step we search for the maximum and minimum concentrations, and define
a contrast parameter $(c_{\mathrm{max}}-c_{\mathrm{min}})/2c$. This parameter
indicates the visibility of the developing pattern.
Another quantity of interest is the pair correlation function 
\begin{equation}
G(\vec{r},t)=\left\langle\frac{1}{V}{\int}(c(\vec{r},t)- c)(  
c(\vec{r}+\vec{r}^\prime,t)-c)d^2\vec{r^\prime}\right\rangle.
\end{equation} 
The average domain width $R_g$ can be calculated from the first zero of the normalized 
pair correlation function $\Tilde{G}(\vec{r},t) = G(\vec{r},t)/G(0,t)$.  
The upper panel of Fig.~\ref{contrast} shows the early time variation of
the contrast parameter and the domain width $R_g$ as a function of the 
scaled time. The average domain width after a rapid initial increase
saturates at this temperature to a value $\sim  4.5\lambda$. 

\section{Summary}
We have studied the electric-field-induced phase separation of a
dielectric binary fluid using a coarse-grained free energy functional. 
The free energy investigated is guided by experiments on field-driven
instabilities in ferro-colloids, and shares features of a 
Ginzburg-Landau free energy that describes the phase separation of a
binary mixture. Certain other features of ferro-particles which
would require a more complicated order parameter have been omitted,
\textit{e.g.} the orientational ordering arising from the magnetic
dipolar interaction. 

The first order co-existence curve of the model clearly shows
how a spinodal decomposition can be driven by an electric field
in colloidal systems. Analytical estimates of the critical field are
consistent with the numerical results. Deep quenches show a typical
labyrinthine pattern formation with characteristic length and time
scales set by the free energy.  

The phenomenology used here is most accurate at low particle
concentrations. In particular, the entropy is essentially that
of free particles, modified to take interactions into account
\textit{via} the phenomenological correlation volume $v$. The
resulting entropy is at least qualitatively correct at low
concentrations but not high. Experiments with low overall concentrations
should be described with reasonably accuracy by this model.

\begin{acknowledgments}
We thank Geoff Canright and John Evans for helpful conversations.
We acknowledge support from the NSF through grants DMR99-72683 (MDJ),
DMR00-72901  (WL), and NSF-NIRT (ENG/ECS and CISE/EIA) grant
0103587 (AB and WL).
\end{acknowledgments}

\pagebreak

\end{document}